\documentclass{amsart}
\usepackage{graphicx} 
\usepackage{color}
\usepackage{caption}
\usepackage{subcaption}
\usepackage{todonotes}
\usepackage{hyperref}
\usepackage{soul}
\usepackage{siunitx}
\usepackage{amsaddr}

\newtheorem{thm}{Theorem}
\newtheorem{cor}{Corollary}
\newtheorem{prop}{Proposition}

\title{Asymmetric relaxations through the lens of information geometry}

\author{Alessandro Bravetti, Miguel Ángel García Ariza, Pablo Padilla}
\address{
Instituto de Investigaciones en Matemáticas Aplicadas y en Sistemas, Universidad Nacional Autónoma de Mexico, A. P. 70543, Ciudad de Mexico 04510, Mexico
}
\email{miguel.garcia@iimas.unam.mx}
\begin{document}

\begin{abstract}
We frame Newton's Law of Cooling as a gradient flow within the context of information geometry. 
This connects it to a thermodynamic uncertainty relation and the Horse-Carrot Theorem, 
and reveals novel instances of asymmetric relaxations in endoreversible processes. 
We present a general criterion for predicting asymmetries using the Amari-Chentsov tensor, 
applicable to classical and quantum thermodynamics. 
Examples include faster cooling of quantum ideal gases and relaxations that resemble the Mpemba effect in classical ideal gases. 
\end{abstract} 

\maketitle

\section{Introduction}

A system in contact with a heat bath undergoes relaxation, 
ultimately reaching an equilibrium state with the same temperature as the bath. 
Nevertheless, certain conditions lead to a faster relaxation process when the system is initially colder than the bath, 
a phenomenon called \textit{asymmetric relaxation} \cite{lapolla_faster_2020}.

To explain, 
consider two systems initially in equilibrium, 
one cooler and one hotter than a  bath with temperature $T_q$. 
After an instantaneous temperature quench, 
the cooler system relaxes faster, 
given \textit{thermodynamically equidistant} initial temperatures \cite{lapolla_faster_2020,ibanez_heating_2024}. 
Here, distance is quantified by the relative entropy of the system distributions. 
Faster relaxation implies that
the relative entropy between the cooler system and the target state is always lower than that of the hotter system. 
This measure of distance, 
also known as \textit{Kullback-Leibler divergence} or \textit{generalized free energy}~\cite{vaikuntanathan_dissipation_2009}, 
is physically meaningful, 
since it represents the difference between the current value of the thermodynamic potential and its equilibrium value. 
Consequently, 
it serves as an (asymmetric) measure for the distance to equilibrium.

This situation applies to systems
characterized by dynamics unfolding in a quadratic potential and governed by an overdamped Langevin equation \cite{lapolla_faster_2020,ibanez_heating_2024}. 
While extending to systems driven from equilibrium with a linear drift \cite{dieball_asymmetric_2023}, 
it still lacks universal applicability \cite{meibohm2021relaxation,van2021toward}.

Here, 
we study asymmetries in relaxations that follow a generalization of Newton's Law of Cooling.
To this end, 
we initially focus on systems that undergo \textit{endoreversible} relaxation, 
where each intermediate stage constitutes an internal equilibrium state, 
and irreversibilities manifest only at the system-bath boundaries~\cite{rubin_optimal_1979,hoffmann_introduction_2008}. 
This ensures a well-defined notion of temperature throughout the process, 
making it possible to analyze relaxations governed by 
\begin{equation}\label{eq:NLC}
    \dot T=-\lambda(T-T_q).
\end{equation}
We begin by using information geometry to describe the latter as a well-known gradient flow \cite{fujiwara_gradient_1995,boumukigradient2016,wada_eikonal_2021,wada_huygens_2023,wada_hamiltonian_2024}.
Then we apply the gradient flow structure as our key tool to derive the main results. 
This approach allows us to extend our analysis to a wider range of relaxations, going well beyond Newton's original law,
including systems out of equilibrium where temperature is not defined throughout the process.

Our work is motivated by the asymmetric relaxation of Gaussian chains,
whose average potential energy and entropy during the process
resemble ideal gas equations \cite{lapolla_faster_2020},
and the relaxation has the same form of Eq.~\eqref{eq:NLC}. 
 
{We start by introducing the information-geometric structure of the space of thermodynamics in Sect.~\ref{sec:DG}. 
This allows us to depict Eq.~\eqref{eq:NLC} as a Riemannian gradient flow in Sect.~\ref{sec:GF}.
We consider the case of closed ideal gases in Sect.~\ref{sec:IG}, 
revealing faster warming than cooling. 
In Sect.~\ref{sec:GR}, 
we demonstrate that this asymmetry transcends specific systems, 
depending only on the gradient flow nature of the relaxation process and the geometric structure of the space of states. 
Sects. \ref{sec:QG} and \ref{sec:Mpemba} illustrate the result's utility, 
exhibiting asymmetry reversal in rigid gases, 
quantum and classical,
and a phenomenon akin to the Mpemba effect in ideal gases. 
Finally, Sect.~\ref{sec:C} concludes with remarks and future perspectives.
We include detailed calculations that lead to the results of Sects. \ref{sec:IG}, 
\ref{sec:QG}, 
and \ref{sec:Mpemba} in appendices. }

\section{The information geometry of thermodynamics}\label{sec:DG}
{
The set of equilibrium states of thermodynamic systems is endowed with a geometric structure consisting of a Riemannian metric $g$,
and two so-called \textit{conjugate connections} $\nabla$ and $\nabla^*$, satisfying
\begin{equation}
      X[g(Y,Z)]=g(\nabla_XY,Z)+g(Y,\nabla^*_{X}Z),\label{eq:dfe}  
\end{equation}
for any vector fields $X$, $Y$, and $Z$ ~\cite{nielsen_elementary_2020}.
This structure,
usually referred to as \textit{dual structure},
generalizes Riemannian geometry in the sense that when these two connections coincide,
they are precisely the Levi-Civita (Riemannian) connection of $g$.
}

{
Let us construct this dual structure explicitly.
According to the Principle of Maximum Entropy,
its equilibrium states are described by probability distributions that maximize Shannon's entropy,
\begin{equation}
    S[p]=-\int_Xp(x)\log p(x)\mathrm{d}x,
\end{equation}
while keeping certain functions $F^i(x)$ constant in average.
The resulting distributions are known as Gibbs' distributions,
and have the following form:
\begin{equation}\label{eq:Gibbs}
    p(x;\eta)=\exp\{F^i(x)\eta_ i-\psi(\eta)\},
\end{equation}
where the $\eta_i$ are Lagrange multipliers,
and we use Einstein's summation convention.
For instance,
in the case of a fluid in a closed container,
the $\eta_i$ represent the inverse of temperature $T$ and the negative pressure $P$ per unit temperature,
respectively (see, for instance, \cite{reichl_modern_1998}).
We will discuss the physical meaning of $\psi$ later.
%\alescomment{Volume is really one of the $F^i$?}
}

{
The parameters $\eta$ in Eq. \eqref{eq:Gibbs} allow us to identify equilibrium states with points on an $n$-dimensional manifold,
which is precisely the \textit{manifold of equilibrium states} of the system.
If we assume that the $F^i$ are independent,
it is not hard to see that $\psi$ is strictly convex.
This means that its Hessian matrix is positive definite,
and therefore defines a metric tensor $g$ on the manifold of states called \textit{thermodynamic metric} 
(more in general, to any family of probability distributions satisfying some regularity conditions one can associate a metric called
the \textit{Fisher metric}; for the case at hand, i.e.~the Gibbs or exponential family, one precisely 
recovers the thermodynamic metric as introduced here).
}

{
To define $g$ in a coordinate-independent fashion,
we introduce a covariant derivative $\nabla^*$, 
that allows us to write $g=\nabla^*\mathrm{d}\psi$. 
Of course, 
since $g^{ij}=\partial^i\partial^j\psi$ in the $\eta$ coordinates,
the Christoffel symbols of $\nabla^*$ in these coordinates must vanish,
whence it is a \textit{flat} connection. 
The $\eta$ are called \textit{affine coordinates} of $\nabla^*$.
}

{
Another consequence of the convexity of $\psi$ is that the functions $\theta^i:=\partial^i\psi$ are independent. 
Hence, they are an alternative parameterization of equilibrium states. 
A straightforward calculation shows that
\begin{equation*}
    \theta^i=E_\eta[F^i]=:\int_XF^i(x)p(x;\eta)\mathrm{d}x.
\end{equation*}
In the case of a fluid in a closed container,
$\theta^1=U$ represents the macroscopic energy of the system,
known as \textit{internal energy},
and $\theta^2=V$ its volume.
}

{
The $\eta_i$ can also be obtained from the $\theta^j$ as follows. 
Observe that
\begin{equation*}
    \mathrm{d}(\theta^i\mathrm{d}\eta_i)=\frac{\partial\theta^i}{\partial\eta_j}\mathrm{d}\eta_j\wedge\mathrm{d}\eta_i=g^{ij}\mathrm{d}\eta_i\wedge\mathrm{d}\eta_j=0.
\end{equation*}
Then, 
$\theta^i=\partial^i\phi$ at least locally, for some function $\phi$.
From this, 
we can see that $g_{ij}=\partial^2\phi/\partial\theta^i\partial\theta^j=:\partial_i\partial_j\phi$ are the components of $g$ in the $\theta$ coordinates,
and that $g_{ij}g^{jk}=\delta_i^k$.
Like we did before,
we may introduce a connection $\nabla$ whose Christoffel symbols vanish in the $\theta$ coordinates,
so that $g=\nabla\mathrm{d}\phi$.
It is straightforward that $\nabla^*$ and $\nabla$ are dual in the sense of Eq. \eqref{eq:dfe}.
Notice that $\nabla$ is also flat.
}

{
The functions $\psi$ and $\phi$ are related with each other:
\begin{equation*}
    \mathrm{d}\phi=\theta^i\mathrm{d}\eta_i=\mathrm{d}(\eta_i\theta^i)-\eta_i\mathrm{d}\theta^i=\mathrm{d}(\eta_i\theta^i)-\mathrm{d}\psi.
\end{equation*}
So, 
up to a constant (which does not alter $g$),
$\phi=\eta_i\theta^i-\psi$ is the Legendre transform of $\psi$. 
Let us elucidate their physical meaning.
}

%{
Going back to Gibbs' distributions,
we have that 
\begin{equation}\label{eq:phi}
    \phi=\eta_i\theta^i-\psi=\eta_iE_\eta[F^i]-\psi=E_\eta[\eta_iF^i-\psi]=E_\eta[\ell_x]=-S,
\end{equation}
where $\ell_x:=\log p(x;\cdot)$ is the log-likelihood of the Gibbs' distribution. 
In equilibrium thermodynamics, 
the Shannon entropy coincides with the thermodynamic entropy of the system.
Thus, $\phi$ is the negative entropy of the system,
regardless of the ensemble we are considering \cite{bravetti_contact_2015,brody_geometrical_1995}.
Its Legendre transform $\psi$ is sometimes called \textit{Massieu function} \cite{callen_thermodynamics_1985}, 
and corresponds to the negative Gibbs free energy $G$ per unit temperature in the case of a fluid in a closed container.

%{
Notice that the dual structure that we constructed relies on the convex function $\psi$ (or its Legendre transform),
rather than on Gibbs' distributions. 
Physically, 
this means that we can construct a similar structure using thermodynamic potentials,
which are convex functions,
with no need to resort to the statistical-mechanical description of equilibrium systems. 
However,
the geometric structures that arise from different potentials are related to the probability of fluctuations in
the corresponding ensembles and therefore are not necessarily equivalent to each other~\cite{bravetti_thermodynamic_2014}.

%{
Convex potentials also introduce a unique divergence,
known as the \textit{Bregman divergence}~\cite{nielsen_elementary_2020}, 
defined for two states $p$ and $q$ by
\begin{equation}\label{eq:D}
D(p||q):=\phi(p)+\psi(q)-{\theta^i}(p)\eta_i(q).
\end{equation}
Divergences are sufficiently smooth, non-negative two-point
functions that vanish only when the points coincide,
providing an asymmetric notion of distance.
In particular, the Bregman divergence coincides with the
relative entropy (Kullback-Leibler divergence) of Gibbs’
distributions when $\phi=-S$. 
When we take $\phi$ to be the internal energy instead,
the correspondence divergence is the \textit{availability} of the system
(see below).
%that is, the amount of work that can be extracted from the system, given that this is not
%at equilibrium with its surroundings.

This is the notion of distance that we will use to measure asymmetries in endoreversible relaxations.

\section{Newton's Law of Cooling as a gradient flow}\label{sec:GF}
We are going to use the dual structure of the space of equilibrium states to portray Eq. \eqref{eq:NLC} as a gradient flow,
as is done in \cite{fujiwara_gradient_1995}.
Let us assume for now that there is a convex function $\phi$ that makes $T=\eta_1$ an affine coordinate of a flat connection $\nabla^*$. 
Then,  Eq.~\eqref{eq:NLC} is one component of 
\begin{equation}\label{eq:syseta}
    \dot\eta_j=-\lambda(\eta_j-{\eta_j}_q),
\end{equation}
where ${\eta_j}_q:=\eta_j(q)$, 
and $q$ is the equilibrium state reached after relaxation. 
In coordinates $\theta^i$, 
Eq.~\eqref{eq:syseta} reads
\begin{equation*}
        \dot\theta^i=g^{ij}\dot\eta_j=-\lambda g^{ij}(\eta_j-{\eta_j}_q)=-\lambda g^{ij}\frac{\partial}{\partial\theta^j}\left(\phi+\psi_q-\theta^k{\eta_k}_q\right),
\end{equation*}
with $\psi_q:=\psi(q)$.
This is equivalent to
\begin{equation}
    \dot\gamma=-\lambda\operatorname{grad}D^*_q,\label{eq:grad}
\end{equation}
where $D^*_q:=D(\cdot||q)$, $\gamma(t):=(\theta^i(t))$, 
and $\operatorname{grad}$ denotes the Riemannian gradient operator. 
This is our first result.

\begin{thm}\label{thm:main1}
    Newton's Law of Cooling is one component of the gradient flow defined by Eq.~\eqref{eq:grad}.
\end{thm}

From Eq. \eqref{eq:syseta}, 
we may see that the gradient flow \eqref{eq:grad} is a $\nabla^*$-pregeodesic~\cite{fujiwara_gradient_1995},
\textit{i.~e.},
\begin{equation}\label{eq:geo}
    \nabla^*_{\dot\gamma}\dot\gamma=-\lambda\dot\gamma.
\end{equation}
This will be useful in the next section.

Now we show that $T$ is indeed an affine coordinate of a flat connection corresponding to a dual structure.
To this end, we resort to the First Law of Thermodynamics,
expressed by 
\begin{equation*}
    \mathrm{d}U=T\mathrm{d}S+\eta_k\mathrm{d}\theta^k.
\end{equation*}
In the case of a fluid in a closed container, 
the only $\eta$ and $\theta$ represent the negative pressure and its volume,
respectively.
In general,
$U$ is a convex function of $S$ and the $\theta^k$,
and thus defines a dual structure where $T$ and the $\eta_k$ are affine coordinates of a flat connection $\nabla^*$.
The metrics of this structure are known as \textit{Weinhold metrics} \cite{weinhold_metric_1975} and are conformally equivalent to Fisher's~\cite{salamon_relation_1984,bravetti2015conformal}, 
also called \textit{Ruppeiner metrics} in the context of equilibrium thermodynamics \cite{ruppeiner_thermodynamics_1979}.

According to Eq.~\eqref{eq:D}, 
the divergence induced by $U$ is
\begin{equation}\label{eq:Dc}
    D^*_q=U+\psi_q-T_qS-\left(\eta_k\right)_q\theta^k,
\end{equation}
which coincides with the negative \textit{availability} of the system. 
This is the amount of useful work that can be extracted when the system is not in equilibrium with its surroundings \cite{callen_thermodynamics_1985, keenan_availability_1951,landau_statistical_2013,pippard_elements_1964},
and provides a notion of ``distance'' in terms of energy.
Mathematically, 
it is proportional to the Kullback-Leibler divergence $D_\text{KL}$ used in \cite{lapolla_faster_2020} 
to measure distance to equilibrium at every stage of a relaxation process,
this is,
\begin{equation*}
    D_q^*=T_qD_\text{KL}(\cdot||q).
\end{equation*}
The $\Lambda_k$ appearing in the distributions describing Gaussian chains there are  equivalent to the $\eta_k$ defined here.
Consequently,
we find it reasonable to use $D^*_q$ as a measure of distance to the target equilibrium state $q$ in endoreversible relaxations dictated by Eq. \eqref{eq:grad}.

We close this section with two straightforward corollaries that follow from regarding Eq.~\eqref{eq:NLC} as a gradient flow. 
The first one establishes a lower bound on the availability dissipated along an endoreversible relaxation dictated by Eq. \eqref{eq:grad}.

Indeed, 
if $\gamma$ satisfies Eq. \eqref{eq:grad}, then
\begin{equation}\label{eq:dotD}
        -\lambda\dot D^*_q=||\dot\gamma||^2.
\end{equation}

On the other hand,
the amount of availability $A$ dissipated along any process $\gamma$ can be expressed as $\mathrm{d}A=-\mathrm{d}D^*_q$ \cite{salamon_thermodynamic_1983,nulton_quasistatic_1985}.
So,
the total dissipated availability after time $\tau$ is
\begin{equation}\label{eq:TUR}
        \delta A(\tau)=\frac{1}{\lambda}\int_0^\tau ||\dot\gamma||^2\mathrm{d}t\geq \frac{L_\gamma^2}{\lambda\tau},
\end{equation}
with $L_\gamma$ denoting the Riemannian length of $\gamma$, 
and the last result following from the Cauchy-Schwarz inequality.
Eq.~\eqref{eq:TUR} is a thermodynamic uncertainty relation between speed
and thermodynamic cost for processes guided by Eq. \eqref{eq:grad}~\cite{crooks2007measuring,ito2018stochastic}.

Assume now that the system is driven along a prescribed path $\gamma$ defined by a sequence of states $\{q(t)\}_{0\leq t\leq\tau}$ 
to which it relaxes, according to Eq. \eqref{eq:grad}.
Then, 
defining $\epsilon:=1/\lambda$ with $\lambda$ time-dependent (\textit{cf}. \cite{diosi_thermodynamic_1996}),
we have that
\begin{equation}\label{eq:HC}
        \delta A=\bar\epsilon\int_0^\tau||\dot\gamma||^2\mathrm{d}t\geq\frac{\bar\epsilon L_\gamma^2}{\tau}.
\end{equation}
Here, $\bar\epsilon$ denotes the  average of $\epsilon$ with respect to $||\dot\gamma||^2$,
and $\tau$ is the duration of the process.
This is known as the Horse-Carrot Theorem \cite{salamon_thermodynamic_1983,salamon_geometry_1998},
which provides a minimum value for the dissipated availability
during \textit{any} process occurring over a finite time (see also \cite{spirkl_optimal_1995}).
Thus, we have proven the following.

\begin{cor}
    %\textcolor{red}{The uncertainty relation~\eqref{eq:TUR} and the} 
    The Horse-Carrot Theorem~\eqref{eq:HC} holds for any process satisfying
    \begin{equation}\label{eq:ltgradient}
        \dot\gamma=-\lambda(t)\operatorname{grad}D^*_{q(t)}.
    \end{equation}
\end{cor}
%\alescomment{should we mention here that this is true also for the uncertainty relation?}

We note that one of the benefits of the information-geometric approach employed in the proof of this result 
is that it extends the applicability of the Horse-Carrot Theorem to any dynamics given by~\eqref{eq:ltgradient}, 
such as those in machine learning and evolutionary biology~\cite{fujiwara_gradient_1995,harper2009information},
with the meaning of $\delta A$ depending on the context.

\section{Asymmetric relaxation of a closed ideal gas}\label{sec:IG}

We now search for asymmetries in endoreversible relaxations of a closed ideal gas. 
Consider two distinct initial equilibrium states $p^\pm_0$ satisfying $D^*_0:=D^*_q(p^+_0)=D^*_q(p^-_0)$ ~\cite{lapolla_faster_2020},
and assume constant pressure,
so that $q$ and $p^\pm_0$ are determined by their temperatures only,
$T_q$ and $T^\pm_0$. 

We use the equations of state of an ideal gas,~\cite{callen_thermodynamics_1985} 
\begin{subequations}
\begin{align}
    U&=cN_0k_\text{B}T\label{eq:cigU},\\
    PV&=N_0k_\text{B}T\label{eq:cigPV},
\end{align}\label{eq:cig}
\end{subequations}
to obtain (see Appendix \ref{app:IG})
\begin{equation}
    S=N_0 k_\text{B}\left[(c+1)\ln\left(\frac{T}{T_q}\right)-\ln\left(\frac{P}{P_q}\right)\right].\label{eq:Scig}
\end{equation}
At constant $\eta_2=-P$, this yields
\begin{equation}
    \Delta D^*=(c+1)N_0k_\text{B}\left[\Delta T-T_q\ln\left(\frac{T^+}{T^-}\right)\right]\label{eq:Dcig},
\end{equation}
where $\Delta D^*:=D_q^*(\gamma^+)-D_q^*(\gamma^-)$ and $D^*$ is given by Eq.~\eqref{eq:Dc}. 
The curves $\gamma^+$ and $\gamma^-$ are solutions to Eq. \eqref{eq:grad} starting at $p^\pm_0$,
respectively.
The constant $N_0$ represents the number of gas particles, 
$c>0$ is related to the energy equipartition,
and $k_\text{B}$ is Boltzmann's constant. 
 
The difference $\Delta D^*$ is always positive, 
irrespective of the initial states. 
This follows from observing that all the critical points of $\Delta D^*$ are maxima,
and that $\Delta D^*(0)=0$ and $\lim_{t\to\infty} \Delta D^*=0$ 
(see Appendix \ref{app:IG}).

\begin{figure}[h!]
    \centering
    \includegraphics[width=\columnwidth]{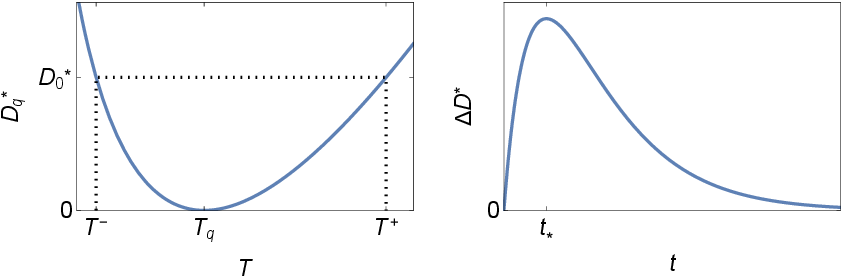}
\caption{Left: {$D^*$} as a function of $T$ for a monoatomic ideal gas ($c=3/2$), in units of $k_\text{B}=1$. 
    We observe that $T^-$ is closer to $T_q$ than $T^+$, 
    meaning that warming up occurs faster at constant pressure. 
    Right: {$\Delta D^*$} as a function of time in units of $k_\text{B}=1$.
    According to the graph,
    the available work is greater when the system is cooling down at every time during the relaxation.
    The difference in available work between the two systems is the greatest at $t=t_*$,
    and it gradually vanishes as the systems relax to equilibrium.}
    \label{fig:IG}
\end{figure}

Thus, we have proved our second main result (see Fig.~\ref{fig:IG}).
\begin{prop}
Given the definition of distance to equilibrium dictated by the thermodynamic availability~\eqref{eq:Dc}, 
a closed ideal gas that relaxes according to Newton's Law~\eqref{eq:NLC} warms up faster than it cools down 
at constant pressure.
\end{prop}

Notice that Eqs.~\eqref{eq:cig} always remain valid during the relaxation, 
provided the process is endoreversible. 
This holds true mathematically, 
even if the initial states fall outside the range of physical validity of Eq.~\eqref{eq:NLC}. 
Remarkably, 
{through an appropriate definition of distance,}
asymmetric relaxations also occur within this regime,
and can be explained {from a geometric standpoint}
(\textit{cf.} \cite{ibanez_heating_2024})
%rather than a far-from-equilibrium one .
%\alescomment{Revisit this sentence?}

% \alescomment{agregar aqu\'i que hemos probado que se vale cerca del equilibrio y que entonces lo que dicen 
% todos que s\'olo es un fen\'omeno lejos del equilibrio no es cierto?}

In view of Eq.~\eqref{eq:NLC} and Fig.~\ref{fig:IG}, 
the asymmetry may seem trivial,
arising from asymmetric initial conditions.
However, 
our non-trivial finding establishes that at any instant of time during the relaxation,
one can extract more useful work from a system cooling down, 
because the availability is always greater in that case.
{Moreover, 
our result extends to situations
where the asymmetry is not predictable from the choice of the initial conditions,
as we will show later.}

\section{The general result}\label{sec:GR}
Let us study the conditions of asymmetric relaxations in a more general setting,
supposing only that the relaxation follows Eq.~\eqref{eq:grad}.
In particular, we do not assume the system has a definite temperature at every instant of time,
nor that it satisfies the standard Newton's Law of Cooling~\eqref{eq:NLC}.

Consider {two solutions} $\gamma^{1,2}$ of Eq.~\eqref{eq:grad}, 
initiated from thermodynamically equidistant conditions $p_0^{1,2}$,
and define {$\Delta D^*:=D_q^*(\gamma^2)-D_q^*(\gamma^1)$.}
Assuming {$\Delta D^*$ }is sufficiently well-behaved, 
asymmetric relaxation occurs when {$\Delta D^*\neq 0$}, for all $t\in(0,\infty)$, 
provided {$\Delta D^*(0)=0$} and {$\lim_{t\to\infty}\Delta D^*=0$.} 
For this,
it suffices that the critical points of {$\Delta D^*$} be only maxima. 

According to Eq.~\eqref{eq:dotD},
$\Delta\dot D^*=0$ if and only if $||\dot\gamma^2||=||\dot\gamma^1||.$ 
{Differentiating $\dot D^*_q$ yields
\begin{equation}\label{eq:ddotD1}
        -\lambda\ddot D_q^*=\dot\gamma[g(\dot\gamma,\dot\gamma)]=\nabla^*_{\dot\gamma}g(\dot\gamma,\dot\gamma)+2g(\nabla^*_{\dot\gamma}\dot\gamma,\dot\gamma)=-C(\dot\gamma,\dot\gamma,\dot\gamma)-2\lambda||\dot\gamma||^2,
\end{equation}
where the last equality follows from Eq.~\eqref{eq:geo}
 and from introducing the \textit{Amari-Chentsov tensor} $C:=-\nabla^*g$}~\cite{amari_information_2016,calin_geometric_2014}.

Evaluating $\Delta \ddot D^*$ at the time $t=t_*$ when $||\dot\gamma^2||=||\dot\gamma^1||$, 
and substituting into the last equation, we derive 
\begin{equation*}
    \lambda\Delta\ddot D^*\vert_{t_*}=C(\dot\gamma^2,\dot\gamma^2,\dot\gamma^2)\vert_{t_*}-C(\dot\gamma^1,\dot\gamma^1,\dot\gamma^1)\vert_{t_*}.\label{eq:main}
\end{equation*}

We have thus arrived at a general characterization of asymmetric relaxations.

\begin{thm}\label{thm:main}
    For two 
    %endoreversible 
    relaxation processes $\gamma^{1,2}$ 
    %of a thermodynamic system 
    satisfying Eq.~\eqref{eq:grad},
    if $C(\dot\gamma^1,\dot\gamma^1,\dot\gamma^1)\geq C(\dot\gamma^2,\dot\gamma^2,\dot\gamma^2)$ 
    whenever $||\dot\gamma^2||=||\dot\gamma^1||$,
    then the relaxation along $\gamma^1$ occurs faster than that along~$\gamma^2$.
\end{thm}

It is important to remark that in the general case all the $\eta$
may vary freely on $\gamma^{1,2}$,
unlike the previous case of isobaric relaxations,
which we now review  in the light of the above theorem. 

In the $\eta$ coordinates, the components of the Amari-Chentsov tensor 
are~\cite{nielsen_elementary_2020}
\begin{equation*}
    C^{ijk}=-\dfrac{\partial^3\psi}{\partial\eta_i\partial \eta_j\partial\eta_k}.
\end{equation*}
At constant pressure,
\begin{equation}
    C=-\dfrac{\partial^3\psi}{\partial T^3}\mathrm{d}T^3.\label{eq:ACpconst}
\end{equation}

Applying Eqs.~\eqref{eq:phi}, \eqref{eq:cig}, and \eqref{eq:Scig}, we have that
$C^{111}=(c+1)N_0k_\text{B}/T^2$, which yields 
\begin{equation*}
        \lambda\Delta \ddot D^*|_{t=t_*}=(c+1)N_0 k_\text{B}\left.\left(\frac{\left(\dot T^+\right)^3}{\left(T^+\right)^2}-\frac{\left(\dot T^-\right)^3}{\left(T^-\right)^2}\right)\right\vert_{t=t_*}<0,
\end{equation*}        
confirming the established result. 

One might wonder about the advantage of computing the Amari-Chentsov tensor over directly calculating $\Delta D^*$ as a function of time. 
The key benefit is that $\Delta D^*$ cannot be obtained analytically, 
even for an ideal gas. 
The complication arises from the choice of initial conditions, 
which must satisfy $D^*_q(p_0^+)=D^*_q(p_0^-)$. 
As the equations of state become more complex, 
like in the next example, 
further difficulties arise. 
In contrast, 
the computation of $C$ is always analytically feasible.

{\section{An example of cooling down occuring faster than warming up}\label{sec:QG}}
We now extend the analysis of the Amari-Chentsov tensor to quantum ideal gases,
illustrating that Theorem~\ref{thm:main}
relies only on the thermodynamic description of the system,
regardless of its classical or quantum nature.
We obtain a rather surprising result: the asymmetry here is opposite, 
cooling down occurs faster than warming up.

To begin, we revisit the equations of state for Fermi-Dirac and Bose-Einstein gases. 
The internal energy $U_\pm$ and particle number $N_\pm$ are \cite{callen_thermodynamics_1985,reichl_modern_1998,pessoa_information_2021}
\begin{subequations}\label{eq:QG}
    \begin{align}
        U_\pm=&\mp\kappa\Gamma(a+2)(k_\text{B} T)^{a+2}\mathrm{Li}_{a+2}(\mp e^{\mu/k_\text{B} T}),\\
        N_\pm=&\mp\kappa\Gamma(a+1)(k_\text{B} T)^{a+1}\mathrm{Li}_{a+1}(\mp e^{\mu/k_\text{B} T})\label{eq:QGN},
    \end{align}
\end{subequations}
where the $+$ ($-$) sign corresponds to fermionic (bosonic) quantities. 
Here, 
$\Gamma$ denotes the Gamma function, 
and $\mathrm{Li}_a$ represents the polylogarithm of order $a\geq1/2$.
The constants $\kappa$ and $a$ are system dependent \cite{pessoa_information_2021}, 
and $\mu$ denotes the chemical potential of the gas.

In this case, the relevant thermodynamic potential is 
\begin{equation*}
    \psi(T,\mu)=\mp\kappa\Gamma(a+1)(k_\text{B} T)^{a+2}\mathrm{Li}_{a+2}(\mp\xi),
\end{equation*}
with $\xi:=\operatorname{exp}[\mu/k_\text{B} T]$ representing the \textit{fugacity} of the gas~\cite{callen_thermodynamics_1985,reichl_modern_1998,pessoa_information_2021}.
For simplicity, we consider relaxations at constant~$\mu$. 
According to Eq.~\eqref{eq:ACpconst},
the only necessary component of $C$ is {(see Appendix \ref{app:QG})}
%\alescomment{I added the reference to this appendix. Is it correct?}
\begin{equation}\label{eq:CQGisomu}
    \begin{split}
        C^{111}_\pm=&\mp\kappa  k_\text{B}^{a-1} T^{a-4} \Gamma (a+1) \left(\mu ^3 \text{Li}_{a-1}\left(\mp \xi\right)\right.\\
       &  + a k_\text{B} T \left\{(a+1) k_\text{B} T \left[3 \mu  \text{Li}_{a+1}\left(\mp \xi\right)\right.\right.\\
       & \left.\left.\left.-(a+2) k_\text{B} T \text{Li}_{a+2}\left(\mp \xi\right)\right]-3 \mu ^2 \text{Li}_a\left(\mp \xi\right)\right\}\right).
    \end{split}
\end{equation}
For our systems of interest, 
all polylogarithmic functions are negative in the fermionic case, 
and positive in the bosonic one, provided that $T>0$.
Besides, if $\mu<0$, $C^{111}_\pm<0$. 
This holds for bosons and for fermions away from the quantum regime ($T\to\infty$ and $\xi\to0$) \cite{reichl_modern_1998}.
Since $\dot T^->0>\dot T^+$,
we have that $C^{111}_\pm(\gamma^+(t))\left(\dot T^+\right)^3>0>
C^{111}_\pm(\gamma^-(t))\left(\dot T^-\right)^3$ for all $t>0.$

\begin{figure}[h!]
\centering
\includegraphics[width=\columnwidth]{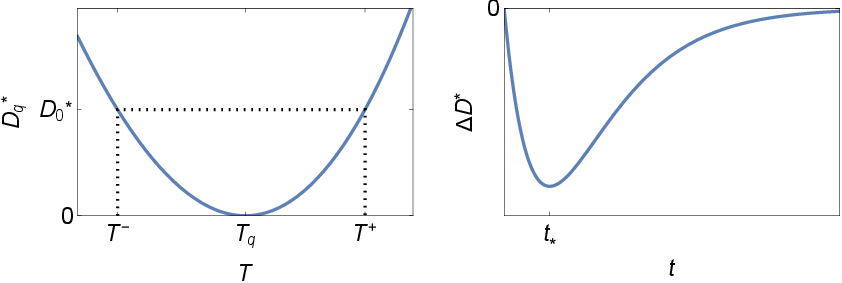}
\caption{Left: $D^*$ as a function of $T$ for bosons, in units of $\kappa=1$ and $k_\text{B}=1$.
    Since $T^+$ is closer to $T_q$ than $T^-$, cooling down occurs faster (\textit{cf.} Fig. \ref{fig:IG}). 
    Right: $\Delta D^*$ for bosons as a function of time, in units of $\kappa=1$ and $k_\text{B}=1$. 
    The shapes of $D^*$ and $\Delta D^*$ agree with the fact that $C(\dot\gamma^+,\dot\gamma^+,\dot\gamma^+)\geq C(\dot\gamma^-,\dot\gamma^-,\dot\gamma^-).$}
    \label{fig:BE} 
\end{figure}

Thus, we have proved the following (see Fig.~\ref{fig:BE}).
\begin{prop}
At constant $\mu$, 
ideal Bose and Fermi gases (away from the quantum regime) cool down faster.
\end{prop}

%\miguelcomment{Aquí o abajo del párrafo de abajo hay que decir algo de por qué en el límite clásico no se voltea la asimetría, pues lo mencionó el primer árbitro en su punto 3.}

{The reversal in the asymmetry during this relaxation process is not a quantum effect. 
We obtain the same result when we consider classical ideal gases enclosed in a rigid container.
The corresponding equations of state are
\begin{subequations}\label{eq:ClassicalRigid}
    \begin{align}
        U&\propto ck_\text{B}T^{c+1}\xi,\\
        N&\propto T^c\xi,
    \end{align}
\end{subequations}
with $\mu<0$. 
The $C^{111}$ component of fermionic and bosonic gases reduce to their classical counterpart in the classical limit given by $T\to\infty$ 
(see Appendix \ref{app:QG}).
}

{
Finally,
we remark that finding $\Delta D^*$} analytically as a function of time is impossible when dealing with quantum gases. 
However, we can determine the asymmetry in the relaxation by means of the explicit calculation of the Amari-Chentsov tensor.
This is also possible in processes involving more than one parameter,
as we illustrate below.

\section{Non-isobaric relaxation of ideal gases: a resemblance to the Mpemba effect}\label{sec:Mpemba}
{
So far, 
we have considered relaxations where only the temperature of the system changes. 
This framework might lead us to believe that asymmetries arise from asymmetric initial conditions,
since $|T_0^--T_q|<|T_0^+-T_q|$ always. 
While this is certainly true in the scenarios we have analyzed to this point,
it is no longer the case when more parameters change during the relaxation. 
In this general situation,
the Euclidean distance between the initial and final states is not meaningful from the physical or geometrical point of view,
and cannot be used to predict asymmetries.
}

{
To illustrate this,}
we compare the \emph{non-isobaric} cooling down of an ideal gas 
to an isobaric cooling down from a higher temperature,
showing that an adequate choice of the initial states  
results in a process resembling the well-known Mpemba effect,
where a hot system freezes faster than a cooler one \cite{mpemba_cool_1969,auerbach_supercooling_1995,bechhoefer_fresh_2021}.
{In this case,
the initial temperature difference is not enough to predict asymmetries,
as we show.}

Consider one mole ($N_0 k_\text{B} \approx\qty{1}{J.K^{-1}}$) of a monoatomic ideal gas ($c=3/2$) prepared at two thermodynamically equidistant states 
$p_0^{1,2}$, both relaxing endoreversibly to a state defined by $T_q=\qty{275}{K}$ and $P_q=\qty{100}{\kilo\pascal}$.
Relaxation from $p_0^1$ is isobaric, with initial temperature 
$T_0^1=\qty{375}{K}$, whereas relaxation from $p_0^2$ has initial temperature 
$T_0^2=\qty{300}{K}$ and pressure $P_0^2=\qty{189.487}{\kilo\pascal}$.

\begin{figure}[h!]
\centering
\includegraphics[width=\columnwidth]{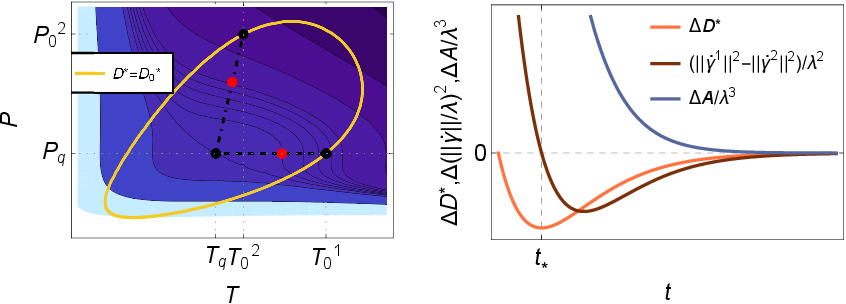}
\caption{Left: Contour plot (dark to light) of $A:=-\lambda^3C(\operatorname{grad}D_q^*,\operatorname{grad}D_q^*,\operatorname{grad}D_q^*)$
    for a monoatomic ideal gas. 
    At $t=t_*$ (red points), 
    $A$ is greater for $\gamma^1$ (dashed) than for $\gamma^2$ (dot-dashed), 
    whence the relaxation along the former is faster. 
    Right: $\Delta D^*$ (orange), $\Delta(||\dot\gamma||/\lambda)^2$ (brown), and $\Delta A/\lambda^3$ (blue) along
    $\gamma^1$ and $\gamma^2$.
    The minimum of $\Delta D^*$ is attained when the relaxation speeds are equal.
    Observe that the two relaxations converge to $q$ at different speeds. 
    Since $\Delta A>0$ at $t=t_*$,
    relaxation along $\gamma^1$ is faster,
    which agrees with $\Delta D^*<0$.
   }
    \label{fig:QuasiMpemba}
\end{figure}

Let $\lambda=\qty{1}{\second^{-1}}$. 
Then, $\Delta D^*$ has a critical point
at $t_*\approx \qty{0.5}{\second}$,
and $C(\dot\gamma^1,\dot\gamma^1,\dot\gamma^1)|_{t_*}>C(\dot\gamma^2,\dot\gamma^2,\dot\gamma^2)|_{t_*}$ {(see Appendix \ref{app:Mpemba})}. 
Hence, the relaxation along $\gamma^1$ is faster 
than that along $\gamma^2$ (see Fig.~\ref{fig:QuasiMpemba}), according to Theorem \ref{thm:main}.

We have thusly revealed a phenomenon similar to the Mpemba effect in ideal gases,
which brings to the surface the underlying geometric aspects of asymmetries in relaxations dictated by Eq.~\eqref{eq:NLC}.
This result is,
of course,
not exactly the same as the Mpemba effect, 
because the systems are not initially identical up to their temperatures {and no phase transition
(e.g.~freezing) is considered. }
In this case, 
the systems are initially identical only in terms of their availability. 

Like in the example of quantum gases, 
$\Delta D^*$ cannot be computed analytically.
However, a straightforward calculation of the Amari-Chentsov tensor yields {(see Appendix \ref{app:Mpemba})}
\begin{equation}
    C=\frac{(c+1)k_\text{B}N}{T^2}\mathrm{d}T^3-\frac{k_\text{B}N}{P^2}\mathrm{d}T\mathrm{d}P^2+2\frac{k_\text{B}N T}{P^3}\mathrm{d}P^3\,,\label{eq:AChIG}
\end{equation}
which can be readily used to analyze the asymmetry in the relaxation.

\section{Concluding remarks}\label{sec:C}
We generalized Newton's Law of cooling using information geometry, 
portraying it as one component of a gradient flow. 
This led us to derive a thermodynamic uncertainty relation between speed and cost, 
as well as the Horse-Carrot Theorem. 
Our approach allows for the study of asymmetric relaxations even near equilibrium, 
suggesting that,
{mathematically,
they can be explained using geometry
in this regime.}
%and that they are not exclusively a result of far-from-equilibrium thermodynamics.

Going beyond Newton's law and considering general gradient flow relaxations,
our results indicate that the Amari-Chentsov tensor is crucial for understanding asymmetries, 
at least where Eq. \eqref{eq:NLC} holds. 
This not only gives it practical physical significance but also underscores its value in the 
geometric description of any dynamics governed by Eq.~\eqref{eq:grad}.
However, its role in more general settings remains an open question.

As with \textit{thermal kinematics} \cite{ibanez_heating_2024}, 
our method is based on information geometry. 
In both cases, 
this provides a geometric understanding of asymmetric relaxations in different regimes. 
Exploring the connection between these approaches offers perspectives for future research.

Through examples, 
we demonstrated the applicability of our results to various thermodynamic systems, 
both classical and quantum. 
Our examination of {rigid systems in Section~\ref{sec:QG} 
reveals an opposite asymmetry compared to the standard one,
both for 
quantum and classical gases.
This} shows the universality of our approach and may offer insights on relaxations in other systems ranging from paramagnetic solids to black holes,
and even others beyond the scope of thermodynamics, 
like machine learning and evolutionary biology~\cite{fujiwara_gradient_1995,harper2009information,amari1998natural,raskutti2015information}.

{It is important to remark that the description of asymmetric relaxations that we follow here relies on a 
specific notion of distance to equilibrium, 
namely, 
the thermodynamic availability of the system.
This choice on the one hand is physically grounded on the fact that it is the most natural choice in terms of energetic (extractable work) analysis.
Besides, 
it coincides with typical choices already used in the literature of asymmetric relaxations if extended to far-from-equilibrium contexts.
However, 
it turns out that this is just one of the many possible choices of a divergence function on the manifold of states.
Describing asymmetric relaxations independently from how distances are measured is an interesting open problem \cite{PhysRevResearch.3.043108}.
It is directly related to} the question of whether the Mpemba effect can be effectively addressed from an information-geometric perspective \cite{vanvu2024thermomajorizationmpembaeffectunification},
presenting another direction for future research.

\appendix
\section{Detailed calculations of the divergence for closed ideal gases}\label{app:IG}
We present here the detailed calculations that lead to our results in Sect. \ref{sec:IG}. 

Let us start with the derivation of $\Delta D^*$. 
We plug Eqs. \eqref{eq:cig} into the First Law of Thermodynamics, 
$\mathrm{d}U=T\mathrm{d}S-P\mathrm{d}V$. 
Then,
we solve for $\mathrm{d}S$,
obtaining
\begin{equation*}
    \mathrm{d}S=N_0k_\text{B}\left(c\frac{\mathrm{d}T}{T}+\frac{\mathrm{d}V}{V}\right).
\end{equation*}
Integrating yields
\begin{equation*}
    S=N_0k_\text{B}\ln\left[\left(\frac{T}{T_q}\right)^c\frac{V}{V_q}\right].
\end{equation*}
Solving Eq.\eqref{eq:cigPV} for $V$ and substituting in the last expression yields Eq. \eqref{eq:Scig}.

To obtain $D_q^*$, 
we substitute Eqs. \eqref{eq:cig} and \eqref{eq:Scig} in Eq. \eqref{eq:Dc}:
\begin{equation*}
    \begin{split}
        D_q^*&=cN_0k_\text{B}T+\psi_q-T_q N_0k_\text{B}\left[(c+1)\ln\left(\frac{T}{T_q}\right)-\ln\left(\frac{P}{P_q}\right)\right]+P_qV\\
        &=N_0k_\text{B}T\left(c+\frac{P_q}{P}\right)+\psi_q-T_q N_0k_\text{B}\left[(c+1)\ln\left(\frac{T}{T_q}\right)-\ln\left(\frac{P}{P_q}\right)\right].
    \end{split}
\end{equation*}
Upon letting $P=P_q$,
we can readily verify Eq. \eqref{eq:Dcig}.

We differentiate this last equation, 
obtaining
\begin{equation*}
    \Delta\dot D^*=(c+1)N_0k_\text{B}\left[\Delta\dot T-T_q\left(\frac{\dot T^+}{T^+}-\frac{\dot T^-}{T^-}\right)\right].
\end{equation*}
Substituting Eq. \eqref{eq:NLC},
\begin{equation*}
    \Delta\dot D^*= -\lambda(c+1)N_0k_\text{B}\Delta T\left(1-\frac{T_q^2}{T^+T^-}\right).
\end{equation*}

Then,
$\Delta D^*$ attains a critical value at $t:=t_*$, 
which satisfies
\begin{equation*}
    T^+T^-\vert_{t=t_*}=T_q^2.
\end{equation*}

To determine the nature of the critical value, 
let us differentiate $\Delta\dot D$ and evaluate in $t=t_*$:
\begin{equation*}
    \begin{split}
        \Delta\ddot D\vert_{t=t_*}&=\left. -\lambda(c+1)N_0k_\text{B}\left[\Delta \dot T\left(1-\frac{T_q^2}{T^+T^-}\right)-\frac{T_q^2\Delta T}{(T^+T^-)^2}\left(\dot T^+T^-+T^+\dot T^-\right)\right]\right\vert_{t=t_*}\\
        & = \lambda^2(c+1)N_0k_\text{B}\left\{\Delta T\vert_{t=t_*}\left.\left(1-\frac{T_q^2}{T^+T^-}\right)\right\vert_{t=t_*}\right.\\
        & -\left.\left.\frac{T_q^2\Delta T}{(T^+T^-)^2}\right\vert_{t=t_*}\left[T_q\left(T^++T^-\right)-2T^+T^-\right]\vert_{t=t_*}\right\}\\
        &= -\lambda^2(c+1)N_0k_\text{B}\left[\frac{\Delta T\vert_{t=t_*}}{T_q^2}T_q\left(T^++T^--2T_q\right)\vert_{t=t_*}\right]\\
        &=-\lambda^2(c+1)N_0k_\text{B}\left[\frac{\Delta T\vert_{t=t_*}}{T_q}\left(T^++T^--2\sqrt{T^+T^+}\right)\vert_{t=t_*}\right]\\
        & = -\lambda^2(c+1)N_0k_\text{B}\left.\left[\frac{\Delta T}{T_q}\left(\sqrt{T^+}-\sqrt{T^-}\right)^2\right]\right\vert_{t=t_*},
    \end{split}
\end{equation*}
which is negative because $\lambda>0$ and $\Delta T=T^+-T^->0$ always. 

Since $\Delta D$ is not identically zero, 
it must attain a maximum at $t=t_*$, 
thusly proving our main claim in Sect. \ref{sec:IG}. 

\section{Detailed calculations of the divergence for quantum ideal gases}\label{app:QG}
Following Theorem \ref{thm:main},
the Amari-Chentsov tensor $C$ is the only object we need to determine asymmetric relaxations dictated by Eq. \eqref{eq:grad}.
To compute it,
we require the third derivatives of $\psi$,
the Legendre transform of $U$,
with respect to $T$ and the $\eta$.
For a gas in a rigid container,
the only $\eta$ is the chemical potential $\mu$.
This is because the First Law in this case has the following form:
\begin{equation*}
    \mathrm{d}U=T\mathrm{d}S+\mu\mathrm{d}N.
\end{equation*}

Recall that $\partial^i \psi=\theta^i$.
So, 
$C^{2ij}$ can be obtained directly with the negative second derivatives of Eq. \eqref{eq:QGN} with respect to $T$ and $\mu$.

The remaining independent components $C^{1ij}$ 
are obtained by computing the partial derivatives of the components of $\mathrm{d}S$ obtained from the First Law:
\begin{equation*}
    \mathrm{d}S=\left(\frac{1}{T}\frac{\partial U}{\partial T}-\frac{\mu}{T}\frac{\partial N}{\partial T}\right)\mathrm{d}T+\left(\frac{1}{T}\frac{\partial U}{\partial \mu}-\frac{\mu}{T}\frac{\partial N}{\partial\mu}\right)\mathrm{d}\mu.
\end{equation*}

In the case of a relaxation at constant $\mu$,
or equivalently $\mathrm{d}\mu=0$,
the Amari-Chentsov tensor reduces to 
\begin{equation*}
    C=-\frac{\partial}{\partial T}\left(\frac{1}{T}\frac{\partial U}{\partial T}-\frac{\mu}{T}\frac{\partial N}{\partial T}\right)\mathrm{d}T^3,
\end{equation*}
which yields Eq. \eqref{eq:CQGisomu}.

The equation above also lets us to find $C^{111}$ for classical ideal gases enclosed in rigid containers.
Indeed, 
plugging in the latter Eqs. \eqref{eq:ClassicalRigid},
we obtain
\begin{equation*}
    C^{111}_\text{c}\propto-k_\text{B}^{-2}T^{c-5}\xi\left[c \left(c^2-1\right) k_\text{B}^3 T^3-3\mu(c-1) c k_\text{B}^2T^2+3 \mu^2 (c-1) k_\text{B}T-\mu ^3\right]
\end{equation*}

To see that Eq. \eqref{eq:CQGisomu} reduces to the last expression when $T\to\infty$,
we consider the greatest power of $T$ in both expressions. 
Additionally, 
we recall that $\operatorname{Li}_b(z)\approx z$ when $z\to0$, for all $b$,
obtaining
\begin{equation*}
    C^{111}_{\pm}\stackrel{T\to\infty}{\propto}-T^{a-1}\xi\text{\quad and \quad }C^{111}_\text{c}\stackrel{T\to\infty}{\propto}-T^{c-2}\xi,
\end{equation*}
which coincide if we consider that the container is a box ($a=1/2$) and the gas has three degrees of freedom ($c=3/2$).

\section{Detailed calculations on the nonisobaric relaxation of ideal gases}\label{app:Mpemba}
Given $p_0^1$,
we wish to find $p_0^2$ satisfying 
\begin{equation*}
    D_q^*(p_0^2) =  D_q^*(p_0^1),\,
    ||\dot\gamma^2(t_*)|| = ||\dot\gamma^1(t_*)||,\,\text{ and }
    C(\dot\gamma^1,\dot\gamma^1,\dot\gamma^1)\vert_{t=t_*}>C(\dot\gamma^2,\dot\gamma^2,\dot\gamma^2)\vert_{t=t_*}
\end{equation*}
for some time $t_*$.
The curves $\gamma^{1,2}$ are $\nabla^*$-pregeodesics starting at $p_0^{1,2}$ and converging to $q$.

The first equation can always be solved,
as long as $p_0^1\neq q$. 
This is true because $D^*_q$ is well-behaved and attains a minimum at $q$,
which means that all its level curves except the one belonging to the critical value will be 1-dimensional submanifolds.
In general, 
it is impossible to obtain this solution analytically,
which underscores the usefulness of Theorem \ref{thm:main}.

Finding general conditions that guarantee the existence of $p_0^2$ is rather challenging. 
Thus, 
we opt for fixing $p_0^1$ and tune the system using $t_*$ as a parameter to obtain an acceptable solution.
For simplicity, 
we set $N_0k_\text{B}=1$ and $\lambda=1$.
The values we choose for $q$ and $p_0^1$ resemble typical laboratory conditions,
and so does our choice of $p_0^2$ among the many solutions we obtain.

To better explain Fig. \ref{fig:QuasiMpemba},
we present the calculations that lead to the quantities involved.

To compute $A$ and $||\operatorname{grad}D_q^*||$, 
we need the metric tensor. 
It is given by the Hessian matrix of the Legendre transform of $U$ with respect to $T$ and $-P$,
or equivalently,
by the first derivatives of $S$ and $V$.
This yields
\begin{equation*}
    g=N_0k_\text{B}\left(\frac{c+1}{T}\mathrm{d}T^2-2\frac{\mathrm{d}T\mathrm{d}P}{P}+\frac{T}{P^2}\mathrm{d}P^2\right).
\end{equation*}

We obtain $\operatorname{grad}D_q^*$ by multiplying the inverse of the matrix representation of $g$ with the matrix representation of $\mathrm{d}D_q^*$.
As expected, 
this results in 
\begin{equation*}
    \operatorname{grad}D_q^*=(T-T_q)\frac{\partial}{\partial T}+(P-P_q)\frac{\partial}{\partial P}.
\end{equation*}
Its norm squared is
\begin{equation*}
    ||\operatorname{grad}D_q^*||^2=N_0k_\text{B}\left[\frac{c+1}{T}(T-T_q)^2-2\frac{(T-T_q)(P-P_q)}{P}+\frac{T}{P^2}(P-P_q)^2\right].
\end{equation*}

Analogously,
we get $A$ by writing $T-T_q$ and $P-P_q$ instead of $\mathrm{d}T$ and $\mathrm{d}P$,
respectively,
in Eq. \eqref{eq:AChIG}

The two relaxations we consider in Sect. \ref{sec:Mpemba} are 
\begin{equation*}
    \gamma^1(t)=(275+100e^{-t},100000)\text{ and }\gamma^2(t)=(25 e^{-t}+275,89487.5 e^{-t}+100000).
\end{equation*}
Plugging these into Eq. \eqref{eq:Dcig} and taking the difference gives
\begin{equation*}
    \begin{split}
        \Delta D_q^* (t) =& \frac{1}{0.00404755\, +0.00452303 e^t}+212.5 e^{-t}+687.5 \log \left(e^{-t}+11\right)\\
       & -687.5 \log \left(4 e^{-t}+11\right)-275. \log (-0.894875 \sinh (t)+0.894875 \cosh (t)+1).
    \end{split}
\end{equation*}
The last expression does not say much,
but allows us to verify straightforwardly that $\Delta D_q^*>0$ for all $t\in(0,\infty)$.
The critical point of this function is $t_*=0.511743$. 

Alternatively,
we may obtain the time dependency of $A$ and $||\dot\gamma||$ in an analogous way.
Then, according to Theorem \ref{thm:main},
we solve $||\dot\gamma^1(t_*)||^2=||\dot\gamma^2(t_*)||^2=$ for $t_*$,
and plug into $\Delta A(t_*)$. 
This procedure yields the same $t_*$,
and $\Delta A(t_*)=21.8106$,
which means that the relaxation along $\gamma^1$ is faster.

\color{black}

\bibliographystyle{unsrt}
\bibliography{references}

\end{document}